\documentclass[12pt,titlepage]{article}
\usepackage{bm, amssymb,pifont,cancel, amsmath,comment,color}
\usepackage[dvips]{graphicx}
\makeatletter

\newcommand{\ltsim}{\protect\raisebox{-0.5ex}{$\:\stackrel{\textstyle <}{\sim}\:$}}
\newcommand{\gtsim}{\protect\raisebox{-0.5ex}{$\:\stackrel{\textstyle >}{\sim}\:$}}

\renewcommand{\theequation}{\thesection.\arabic{equation}}
\@addtoreset{equation}{section}
\makeatother
 
\begin{document}

\begin{titlepage}
\null
\begin{flushright}
WU-HEP-12-08
\\

November, 2012
\end{flushright}

\vskip 1.5cm
\begin{center}
\baselineskip 0.8cm
{{\LARGE \bf
Instant uplifted inflation:}\\
\large A solution for a tension between inflation and SUSY breaking scale
}

\lineskip .75em
\vskip 1.5cm

\normalsize

{\large\bf Yusuke Yamada}$\!${\def\thefootnote{\fnsymbol{footnote}}
\footnote[1]{E-mail address: yuusuke-yamada@asagi.waseda.jp}}

\vskip 1.0em

{\small\it Department of Physics, Waseda University, \\ 
Tokyo 169-8555, Japan}

\vspace{12mm}

{\bf Abstract}\\[5mm]
{\parbox{13cm}{\hspace{5mm} \small
The Hubble parameter during an inflationary era must be smaller than the gravitino mass if moduli fields are stabilized by the Kachru-Kallosh-Linde-Trivedi mechanism. This condition represents a difficulty to combine low scale SUSY breaking and high scale inflation. We propose a simple mechanism which can naturally separate the inflation scale from the SUSY breaking scale today. 
}}

\end{center}

\end{titlepage}

\tableofcontents
\vspace{35pt}
\hrule
\section{Introduction}
Cosmological inflation~\cite{Guth:1980zm}  is strongly favored. That is because it can solve some theoretical problems of standard cosmology (e.g. the flatness and the horizon problem). On the other hand, typical slow-roll inflation models provide a source of the primordial density perturbations which are almost scale invariant (for a recent review see Ref.~\cite{Mazumdar:2010sa}). Such density perturbations are favored for a consistency with the cosmic microwave background (CMB) observation. Therefore observational results also support the existence of an inflationary era. 

In particle physics, supersymmetry (SUSY) is a promissing extension beyond the standard model (SM) of elementary particle. SUSY can solve some theoretical problems such as the gauge hierarchy problem and absence of dark matter candidates in the SM.  Observational and experimental data indicate that SUSY must be broken, and SUSY breaking scale should be above the energy scale of the electroweak symmetry breaking. In the minimal supersymmetric standard model (MSSM), the $Z$ boson mass is related to soft SUSY breaking masses of the MSSM Higgs fields. Recently, Higgs boson was discovered at the large hadron collider (LHC) with its mass $m_H\sim 125$GeV. Then, if we consider SUSY as a solution for the gauge hierarchy problem, TeV-scale SUSY breaking models are favored from the viewpoint of naturalness. In addition, MSSM with low scale SUSY breaking provides good candidates for dark matter. Therefore, SUSY models with low scale SUSY breaking are fascinating.

In order to discuss both the cosmological inflation and the (MS)SM together, a self-consistent quantum gravity theory such as superstring theory is required. Superstring theory is the most promising candidate for the theory of quantum gravity which has a possibility to explain results from cosmological observations and collider experiments. Superstring theory predicts six dimensional extra dimension, whose volume and shape are determined by vacuum expectation values (VEVs) of moduli fields. In four dimensional (4D) effective theories, parameters in the (MS)SM such as gauge coupling constants are also determined by those VEVs, and then moduli stabilization is an important issue. 

The Kachru-Kallosh-Linde-Trivedi (KKLT) model~\cite{Kachru:2003aw} is a well known moduli stabilization scenario in superstring, where the stabilization mechanism consists of three steps. First, it is assumed that the dilaton and the complex structure moduli are stabilized through three form fluxes at a high scale. At the second step, the K$\ddot{\rm{a}}$hler modulus dependent term in superpotential is introduced assuming certain non-perturbative effects. Thus, the K$\ddot{\rm{a}}$hler modulus is stabilized by such a superpotential. But, the minimum of the scaler potential is negative valued. At the third step, the anti de Sitter (AdS) vacuum is uplifted to the de Sitter vacuum by a SUSY breaking term with a positive energy.

However, a rather unexpected problem was pointed out in Ref.~\cite{Kallosh:2004yh}. The problem is that in KKLT type models the Hubble parameter $H$ and the gravitino mass $m_{3/2}$ must satisfy a condition $H\ltsim |m_{3/2}|$ to stabilize the moduli during inflation. As a consequence, SUSY models such as $m_{3/2}\sim\mathcal{O}(1)$TeV require that the scale of $H$ during inflation must be below TeV. Then, it is difficult to generate the scalar perturbation consistent with the CMB observations. This problem was pointed out by Kallosh and Linde, which will be called the Kallosh-Linde (KL) problem in this paper. 

The same problem occurs even in moduli inflation models. From the minimalistic point of view, it seems natural to consider the case that moduli fields play a roll of inflaton, and such models were suggested.(For example, see Refs.~\cite{BlancoPillado:2004ns}, \cite{Linde:2007jn}, and \cite{Conlon:2005jm}.) They are successful models from the viewpoint of inflation, however do not realize low scale SUSY breaking models. In Refs.~\cite{Badziak:2008yg} and \cite{Covi:2008cn}, it was shown that the volume modulus which has a K$\ddot{\rm{a}}$hler potential term given by $K=-n\log (T+\bar{T})$ with $0< n\leq 3$ can not have an inflationary de Sitter point without SUSY breaking terms. (We call such moduli fields as volume type moduli.) Then, in typical moduli inflation models, SUSY breaking terms are required to realize an inflationary de Sitter point. Therefore the SUSY breaking scale is related to the inflation scale in the same way as the KL problem. In addition, the moduli have a possibility to overshoot the minimum and to be destabilized after inflation. This is so-called the overshooting problem. This problem also causes a difficulty in moduli inflation models.

On the other hand, low scale SUSY breaking models with the KKLT stabilization have interesting phenomenological aspects. For example, a mixing of the modulus and the anomaly mediation leads to so-called the mirage mediation~\cite{Choi:2005hd} which is a solution of the SUSY little hierarchy problem. And then, the modulus mediated SUSY breaking is fascinating. 

In this paper, we propose a new mechanism to combine high scale inflation with low scale SUSY breaking in which the F-term of the modulus is similar to the original KKLT model. Two important ingredients are required for the realization of our models. One is a SUSY breaking field $Y$ which has a superpotential term $W=\mu_Y^2Ye^{-c_YT}$. This kind of terms generates F-terms varying exponentially in terms of $T$. Then, even if such a F-term makes an inflationary de Sitter point at high scale, the F-term becomes exponentially small as the modulus rolls into its mildly large VEV. The other one is a non-perturbative superpotential term with positive exponents~\cite{Abe:2005rx}. As pointed out in Ref.~\cite{Abe:2005rx}, it can prevent the overshooting problem without fine-tuned initial conditions and parameters. 

This paper is organized as follows. In Sec. 2, we review a reason why it is difficult to combine high scale inflation models with the low energy SUSY breaking. Then, we discuss a model with the two ingredients mentioned above, and show that those ingredients can separate the SUSY breaking scale from the inflation scale in Sec. 3. In Sec. 4, we show concrete models with different types of moduli stabilization. Finally, we conclude in Sec. 5.
\section{Review of the KL problem}
In typical inflation models with moduli, the gravitino mass $m_{3/2}$ today are related to the Hubble parameter during an inflationary era. Therefore, the low scale SUSY breaking models with KKLT stabilization leads to  low scale inflation.  One of the possibility is the case that (a)~moduli are not inflatons. The other one is the case that (b)~moduli are inflatons. 

First, we review the situation (a) which causes the KL problem. To discuss concretely, we consider a simple KKLT type model~\cite{Kachru:2003aw} where the modulus field $T=\sigma +i\alpha$ has the K\"ahler potential and the superpotential as follows:
\begin{eqnarray}
K&=&-3\log (T+\bar{T}),\\
W&=&w_0+A e^{-a T}.
\end{eqnarray}
The F-term scalar potential in 4D supergravity~(SUGRA) can be written in terms of $W$ and $K$ in the following form:
\begin{eqnarray}
V=e^K(D_IW K^{I\bar{J}}D_{\bar{J}}\bar{W}-3|W|^2) \label{eq;scalar potential},
\end{eqnarray}
where $D_IW=\partial_IW+\partial _IK W$, the indices $I, \bar{J}$ denote corresponding chiral superfields $Q^I$ and their conjugates $\bar{Q}^{\bar{J}}$ respectively, and $\partial_I$ represents a derivative with respect to the lowest component of a chiral superfield $Q^I$. 

In this case, the SUSY condition $D_TW=0$ is satisfied at the minimum. Then, a VEV of the scalar potential at the SUSY point is given by (in the Planck unit $M_{\rm{pl}}=1$)
\begin{eqnarray}
\langle V\rangle _{\rm{AdS}}=-3e^{\langle K\rangle }|\langle W\rangle |^2=-3m_{3/2}^2 \label{eq;ads}.
\end{eqnarray}
To vanish the cosmological constant (to be precisely, the value has to be $\Lambda=\mathcal{O}(10^{-120})$), SUSY breaking terms are required\footnote{In the original model~\cite{Kachru:2003aw}, they consider an uplifting term from anti-D3 branes. As shown in Refs.~\cite{Abe:2006xp}, non-vanishing F-terms of chiral superfields can also uplift the potential minimum.} to uplift the minimum:
\begin{eqnarray}
V&=&e^K(D_IW K^{I\bar{J}}D_{\bar{J}}\bar{W}-3|W|^2) +V_{\rm{uplift}},\label{eq;uplifted}\\
V_{\rm{uplift}}&\sim& |3m_{3/2}^2|.\nonumber
\end{eqnarray}

\begin{figure}
\begin{center}
\includegraphics[width=0.5\textwidth]{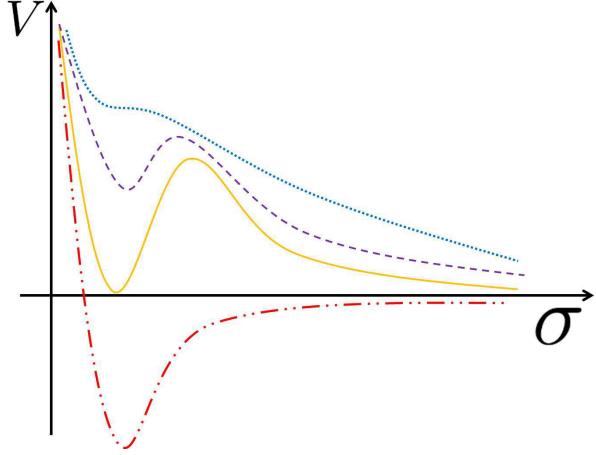}
\caption{An illustration of the F-term potential without uplifting terms drawn by the red dashed-two-dotted line. (See Eq.~(\ref{eq;ads}).) The uplifted potential is drawn by a yellow normal line. (See Eq.~(\ref{eq;uplifted}).) As the inflationary potential becomes larger (see Eq.~(\ref{eq;inflation})), the barrier becomes smaller or disappears. (Drawn by the purple dashed line and blue dotted line respectively.) }
\label{fig;uplift}
\end{center}
\end{figure}

As shown schematically in Fig.~\ref{fig;uplift}, the potential has a barrier after uplifting. The hight of the barrier is approximately given by
\begin{eqnarray}
V_{B}\sim \mathcal{O}(|\langle V_{\rm{AdS}}\rangle |)\sim \mathcal{O}(m_{3/2}^2).
\end{eqnarray}
The scalar potential is generalized to the model with an inflaton field. For simplicity, we do not consider the case that the modulus $T$ and the inflaton $\phi$ have mixing terms in the K$\ddot{\rm{a}}$hler potential and the superpotential. Then, the scalar potential becomes a following form:
\begin{eqnarray}
V&=&e^K(D_{\phi}WK^{\phi \bar{\phi}}D_{\bar{\phi}}\bar{W}+D_TWK^{T\bar{T}}D_{\bar{T}}\bar{W}-3|W|^2)+V_{\rm{uplift}}\nonumber\\
&\sim& \frac{1}{8\sigma^3}(D_TWK^{T\bar{T}}D_{\bar{T}}\bar{W}-3|W(T)|^2)+V_{\rm{uplift}}+\frac{V_{\rm{inf}}(\phi)}{8\sigma^3}, \label{eq;inflation}
\end{eqnarray}
where $V_{\rm{inf}}(\phi) $ denotes the inflaton dependent term and we assume that $V_{\rm{inf}}(\phi) $ vanishes at the minimum. Then we find that the term $\frac{V_{\rm{inf}}(\phi)}{8\sigma^3}$ has a positive value during the inflationary era. That term plays the same roll as the uplifting term $V_{\rm{uplift}}$. As the inflaton potential becomes larger than $|\langle V\rangle _{\rm{AdS}}|$, the scalar potential at the minimum during inflation with respect to the moduli becomes higher and the barrier becomes smaller or disappears. (See Fig.~\ref{fig;uplift}.) To avoid such a destabilization, the Hubble parameter needs to satisfy the condition :
\begin{eqnarray}
H\ltsim m_{3/2}\label{eq;KLproblem}.
\end{eqnarray}
This is the KL problem.

Kallosh and Linde pointed out the KL problem in Ref.~\cite{Kallosh:2004yh}, and suggested a simple solution\footnote{Recently, some alternative models to solve the KL problem were suggested by Refs.~\cite{Kobayashi:2010rx}, \cite{He:2010uk}.}. It is called the  KL model which contains the following alternative superpotenital terms:
\begin{eqnarray}
W_{\rm{KL}}&=&w_0+Ae^{-aT}-Be^{-bT},\label{KL}\\
w_0&=&B\left(\frac{aA}{bB}\right)^{\frac{b}{b-a}}+A\left(\frac{aA}{bB}\right)^{\frac{a}{b-a}}.\label{eq;KLcondition}
\end{eqnarray}
The stationary point satisfies $D_TW=0$. (We denote the value of modulus at this point as $T=T_{\rm{min}}$.) Unlike the original KKLT model, we find that $W=0$ at $T=T_{\rm{min}}$ because of the condition~(\ref{eq;KLcondition}). Then, the minimum is the Minkowski vacuum, and the hight of the potential barrier is not related to the gravitino mass $m_{3/2}$. This model seems simple, however the condition~(\ref{eq;KLcondition}) requires a fine-tuning of the parameter $w_0$\footnote{Some of recent works \cite{Kallosh:2011qk}, \cite{Linde:2011ja} show that the KL model requires a fine-tuning as the same order of a tuning as the original KKLT model if the small constant term $w_0$ is induced by fine-tuned  fluxes. However, Ref. \cite{Abe:2006xi} showed that the ``smallness" of $w_0$ without fine-tunings can be realized if the constant term induced by fluxes is zero and the dilaton $S$ dependent gaugino condensation term provides an effective constant term $\langle Ae^{-a S}\rangle$. Then the smallness of  $w_0$ is not a consequence of fine-tunings. }
. 
Secondly we review problems in the situation~(b). In this case, the general conditions (\ref{eq;Hubble_uplift}) shown later are necessary for a realization of the inflationary de Sitter point. That constraint is originated in the inequality (\ref{eq;ineq}) studied in detail in Refs.~\cite{Badziak:2008yg}, \cite{Covi:2008cn} and shown later. Based on Ref.~\cite{Covi:2008cn}, we discuss about that condition. In the following discussion, we use quantities defined by
\begin{eqnarray}
G&\equiv& K+\log |W|^2,\\
\gamma&\equiv& \frac{V}{3e^G}=\frac{V}{3m_{3/2}^2},\\
f_I&\equiv&\frac{G_I}{\sqrt{G^{J\bar{K}}G_JG_{\bar{K}}}},\\
G^{I\bar{J}}&\equiv&(G_{I\bar{J}})^{-1}=(\partial_I\partial_{\bar{J}}G)^{-1},\\
R_{I\bar{J}K\bar{L}}&\equiv& \partial _I\partial_{\bar{J}} G_{K\bar{L}}-G^{M\bar{N}}\partial_{\bar{J}}G_{M\bar{L}}\partial_I G_{K\bar{N}},\\
\hat{\sigma}(f^I)&\equiv& \frac{2}{3}-R_{I\bar{J}K\bar{L}}f^If^{\bar{J}}f^Kf^{\bar{L}}.
\end{eqnarray}
Here we consider the case in which inflatons do not have canonical kinetic terms. Generalized slow-roll parameters are given by 
\begin{eqnarray}
\epsilon&=& \frac{\nabla_IV G^{I\bar{J}}\nabla_{\bar{J}}V}{V^2},\\
\eta&= &{\rm{minimum \ eigenvalue \ of}} \ {\bf{M}}\\
{\bf{M}}&=& \frac{1}{V} \left( \begin{array}{cc}\nabla ^I\nabla_JV &\nabla ^I\nabla_{\bar{J}}V,\\
\nabla ^{\bar{I}}\nabla_JV & \nabla ^{\bar{I}}\nabla_{\bar{J}}V \end{array} \right) \nonumber
\end{eqnarray}
where $\nabla _I $ is the covariant derivative of the K\"ahler manifold whose metric is given by $G_{I\bar{J}}$, and $\eta$ denotes the minimum eigenvalue of the matrix $\bf{M}$. As a result of general discussions, we find the upper bound on $\eta$: \footnote{One can find the derivation of this relation in Ref.~\cite{Covi:2008cn}.}
\begin{eqnarray}
\eta \leq -\frac{2}{3}+\frac{4\sqrt{\epsilon}}{\sqrt{3(1+\gamma)}}+\frac{\gamma \epsilon}{1+\gamma}+\frac{1+\gamma}{\gamma}\hat{\sigma}(f^I)\sim \frac{1+\gamma}{\gamma}\hat{\sigma}(f^I)-\frac{2}{3}\label{eq;ineq}.
\end{eqnarray}
For achieving successful inflation, $|\eta|$ and $\epsilon$ must be small during inflation. As shown in Ref.~\cite{Covi:2008cn}, we find  $\hat{\sigma}(f^I)\leq 0$ for $K=-3\log(T+\bar{T})$. This leads the relation $\eta \ltsim -\frac{2}{3}$ namely $ |\eta|\gtsim\frac{2}{3}$. In Ref.~\cite{Badziak:2008yg}, it was pointed out that this relation holds in any case for $K=-n\log(T+\bar{T})$ $0< n\leq3$. Therefore, it seems that successful moduli inflation can not be realized. However, this claim does not hold if the scalar potential includes F-terms of other fields~\cite{Badziak:2009eh} or explicit SUSY breaking terms (e.g. anti-D3 brane)~\cite{Badziak:2008yg}. Then, such terms may give inflationary de Sitter points for the scalar potential.  Both of them play a role of  uplifting term like $V_{\rm{uplift}}$ in Eq.~(\ref{eq;uplifted}), and then the height of the inflationary de Sitter point $V_{\rm{inf}}\sim H^2$ satisfies the relation:
\begin{eqnarray}
\mathcal{O}(H^2)\sim\mathcal{O}(V_{\rm{uplift}})\sim\mathcal{O}(m_{3/2}^2).\label{eq;Hubble_uplift}
\end{eqnarray}
This relation is similar to Eq.~(\ref{eq;KLproblem}). Therefore, again we cannot combine high scale inflation with low scale SUSY breaking.

There are some models avoiding the relation~(\ref{eq;Hubble_uplift}). In Refs.~\cite{Badziak:2008yg} and \cite{Badziak:2008gv}, the K$\ddot{\rm{a}}$hler potential term of the volume modulus includes the $\alpha'$-correction, and it changes the value of $\hat{\sigma}(f^I)$. These models can be solutions for a tension between inflation and SUSY breaking scale, however, both of them require fine-tunings of parameters in the superpotential to separate the inflation scale from the SUSY breaking scale as is the case for the KL model. In Ref.~\cite{Conlon:2008cj}, it was suggested that the SUSY breaking scale is much smaller than the inflation scale if the VEV of the volume modulus change into the large volume minimum after inflation\footnote{The K\"ahler potential in this model also contains the $\alpha '$ correction contribution, and then the inflationary de Sitter point is generated.}. Although the way to separate the two scales is interesting, there is the overshooting problem because of an extreme difference between the scale of the inflationary potential and the the barrier at the minimum. Therefore, the large volume model\cite{Conlon:2008cj} requires to choose initial conditions precisely. A simple solution for the overshooting problem is positive exponent terms discussed in Ref.~\cite{Abe:2005rx}. However, such positive exponent terms prevent the volume moduli from obtaining extremely large VEVs, and then it is difficult to combine positive exponent terms with the large volume inflation model.\footnote{We would like to thank Tetsutaro Higaki for pointing out this issue.}
\section{Instant uplifted inflation}
In this section, we propose a new mechanism which can separate inflation scale from SUSY breaking scale. In that model, there are two important ingredients. One is the 
existence of a chiral superfield which has the superpotential term as follows:
\begin{eqnarray}
W_{\rm{uplifton}}=\mu_Y^2Ye^{-c_YT}. \label{eq;uplifton}
\end{eqnarray}
In this paper, we refer to such a field $Y$ as ``uplifton". As mentioned in section 2, a volume-type modulus inflation requires SUSY breaking terms for a realization of an inflationary de Sitter point. The F-term of an uplifton can be source of such an inflationary point. And the VEV of its F-term decreases exponentially with increasing VEV of the modulus $T$. This feature enables a separation of the two scales. The term like Eq.~(\ref{eq;uplifton}) can arise, e.g., from string instanton effects~\cite{Florea:2006si} or anomalous U(1) couplings~\cite{Cvetic:2008mh}. Even in effective theory of simple 5D SUGRA models on $S^1/Z_2$, the factor $e^{-c_YT}$ is always associated with bulk matter fields in the superpotential induced at one of the fixed point, if  bulk matters are charged under the $ Z_2$ odd U(1) gauge vectors~\cite{Abe:2006eg}. 

We consider a model in which the superpotential is given by
\begin{eqnarray}
W=\mu_Y^2Ye^{-c_YT}+Ae^{-aT}-Be^{-bT}.
\end{eqnarray}
In this model, the F-term potential (\ref{eq;scalar potential}) is proportional to terms with negative exponents of $T$. Therefore, as the modulus rolls into the direction of a large VEV, the scale of the scalar potential decreases exponentially that makes the SUSY breaking scale today small enough irrespective of the magnitude of the initial inflationary scale.
This mechanism seems similar to the one in Ref.~\cite{Conlon:2008cj} in which the volume modulus reach the large volume minimum. However, there is a broad distinction between them. In our model, the modulus doesn't have to reach an extremely large VEV though a separation of the two scales can be realized. As we will see in the following, the exponentially decreasing feature is important to combine this separation mechanism with the second important ingredient explained below.

The second key ingredient is the positive exponent term~\cite{Abe:2005rx}. By virtue of the fact that the modulus doesn't need to have an extremely large VEV, we can add positive exponent terms in the superpotential\footnote{Some moduli stabilization models with positive exponent terms are considered in Ref.~\cite{Abe:2005rx} and their application to inflation models can be found in Refs.~\cite{Badziak:2008gv},\cite{Abe:2005rx} and \cite{Kobayashi:2010rx}. } such as
\begin{eqnarray}
W_{\rm{positive}}=\tilde{A}e^{aT}, \label{eq;positive}
\end{eqnarray}
where
$$\tilde{A}=Ae^{-a'\langle S\rangle}.$$
The field $S$ is a heavy modulus which is already stabilized at a higher scale than the cut off scale in our discussion.
Such a positive exponent term can be generated, e.g., if we consider the gauge kinetic function which has a form: 
\begin{eqnarray*}
f=w_SS+w_TT.
\end{eqnarray*}
Then, if gaugino condensation occurs for the SU(N) gauge group with the above gauge kinetic function, the following superpotential term is generated:
\begin{eqnarray}
W=Ae^{-\frac{2\pi}{N}(w_SS+w_TT)}\label{eq;gauge mixing}.
\end{eqnarray}
As mentioned in Ref.~\cite{Abe:2005rx}, the coefficient $w_T$ may have a negative value in some cases (e.g., in heterotic M theory~\cite{Lukas:1997fg}, or magnetized D9-brane~\cite{Cascales:2003zp}). Such gauge kinetic functions also can be realized in 5D SUGRA models where the moduli mixing in gauge kinetic functions are determined by arbitrary coefficients of the cubic polynomials governing the structure of $\mathcal{N}=2$ gauge vector multiplets whose fifth components correspond to moduli $S$ and $T$~\cite{Ceresole:2000jd}.
\begin{figure}[htbp]
\begin{center}
\includegraphics[width=0.8\textwidth]{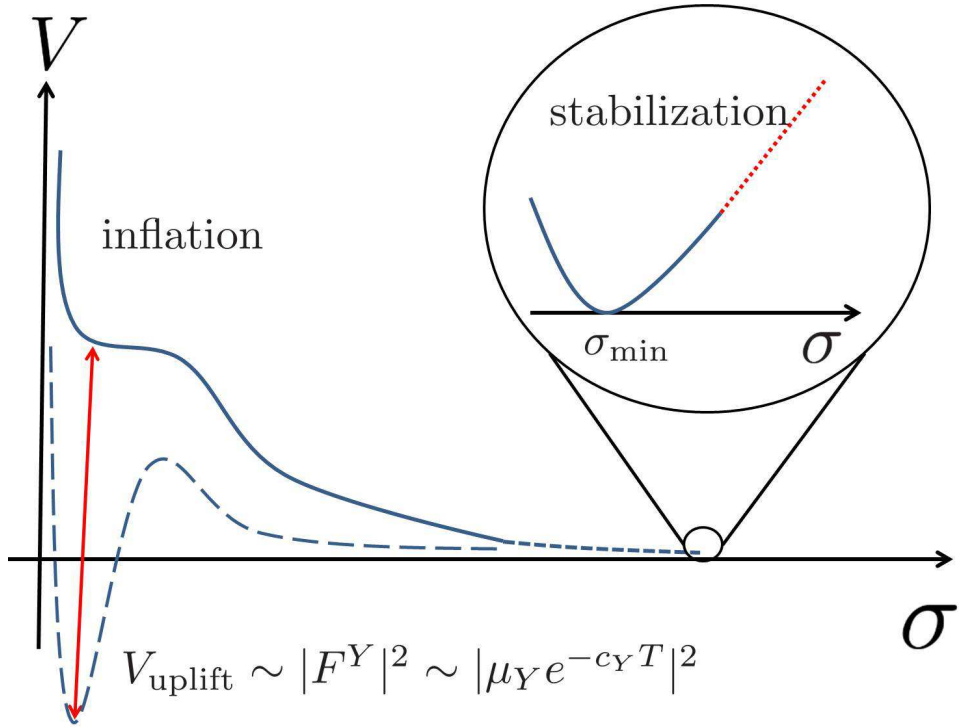}
\caption{An schematic illustration of the instant uplifted inflation scenario explained in Sec.~3. The F-term of the uplifton may yield the inflationary de Sitter point at high scale. The minimum is separated from that point, and the scale near the minimum is much smaller than the inflationary region. The potential includes a positive exponent term, then it will blow up for $\sigma$ larger than $\sigma_{\rm{min}}$. The blowing up feature is drawn by the red dotted line.}
\label{fig;schematic}
\end{center}
\end{figure}

Uplifton and a positive exponent term can realize the separation of the two scales. 
We show the scalar potential which includes the two ingredients in Fig.~\ref{fig;schematic} schematically. To make the mechanisms clear, let's consider a model in which the superpotential is given by
\begin{eqnarray}
W&=&W_{\rm{inf}}+W_{\rm{min}},\\
W_{\rm{inf}}&=&Ae^{-aT}-Be^{-bT}+\mu_Y^2Ye^{-c_YT},\\
W_{\rm{min}}&=&w_0-\tilde{C}e^{cT}+\mu_X^2X.
\end{eqnarray}
We assume following conditions:
\begin{eqnarray}
c_XT_{\rm{inf}}&\sim& aT_{\rm{inf}}\sim bT_{\rm{inf}}\sim\mathcal{O}(1),\\
cT_{\rm{min}}&\sim&\mathcal{O}(1),\\
T_{\rm{min}}&\sim& 10T_{\rm{inf}}, \label{eq;min_inf}\\
|\tilde{C}e^{cT_{\rm{inf}}}|&<&|w_0|\ll |Ae^{-aT _{\rm{inf}}}|, |Be^{-bT _{\rm{inf}}}|, |\mu_Y^2e^{-c_YT_{\rm{inf}}}|,
\end{eqnarray}
where $T_{\rm{inf}}$ denotes the typical VEV of $T$ around the inflationary point, and $T_{\rm{min}}$ denotes the one around the minimum.
Then, the scalar potential around the inflationary de Sitter point is dominated by $W_{\rm{inf}}$. We can represent the scalar potential around the inflationary de Sitter point by
\begin{eqnarray}
V|_{T\sim T_{\rm{inf}}}\sim \frac{1}{8\sigma^3}(\mu_Y^4e^{-2c_Y\sigma}+D_T\hat{W}_{\rm{inf}}K^{T\bar{T}}D_{\bar{T}}\hat{\bar{W}}_{\rm{inf}}-3|\hat{W}_{\rm{inf}}|^2),
\end{eqnarray}
where $\hat{W}\equiv W|_{X=Y=0}$.

The dominating part of the superpotential terms will change however, from $W_{\rm{inf}}$ to $W_{\rm{min}}$ after the moduli rolling down into the minimum. That is because the condition~(\ref{eq;min_inf}) leads $c_XT_{\rm{min}}\sim aT_{\rm{min}}\sim bT_{\rm{min}}\sim\mathcal{O}(10)$, then $|W_{\rm{inf}}|$ becomes an exponentially suppressed value. Therefore, the scalar potential around the minimum is mainly determined by $W_{\rm{min}}$. In this situation, the scalar potential is represented by
\begin{eqnarray}
V|_{T\sim T_{\rm{min}}}\sim \frac{1}{8\sigma^3}(\mu_X^4+D_T\hat{W}_{\rm{min}}K^{T\bar{T}}D_{\bar{T}}\hat{\bar{W}}_{\rm{min}}-3|\hat{W}_{\rm{min}}|^2).
\end{eqnarray}
We don't have to care about the overshooting, hence $\hat{W}_{\rm{min}}$ contains a positive exponent term in the scalar potential. The only required condition for the superpotential $W_{\rm{min}}$ is found that it contains at least a single positive exponent term. Therefore we can consider some models with different types of stabilization potential aside from positive exponent terms. We will see some illustrative models in the next section.
\section{Some illustrative models}
We show three concrete models to realize the mechanism discussed in Sec. 3. Those three models are different with respect to the stabilization potential. In all models, K\"ahler potential $K$, and superpotential terms $W_{\rm{inf}}$ dominating inflation are as follows:
\begin{eqnarray}
K&=&-3\log (T+\bar{T})+|X|^2-\frac{1}{\Lambda ^2}|X|^4+|Y|^2-\frac{1}{\Lambda ^{'2}}|Y|^4,\label{eq;kahler}\\
W&=&W_{\rm{inf}}+W_{\rm{min}},\\
W_{\rm{inf}}&=&Ce^{-cT}-De^{-dT}+\mu _Y^2Ye^{-c_YT}\label{eq;Winf}.
\end{eqnarray}
It is difficult to solve the dynamics of multiple fields simultaneously. Here it is chosen a set of parameters such that the masses of the fields other than the inflaton ($\sigma={\rm{Re}}T$) are heavy, and then we treat following models as single field inflation. In this case, we can analyze the relevant part of  the whole dynamics based on the following effective potential:
\begin{eqnarray}
V_{\rm{eff}}=\frac{1}{8\sigma^3}(\mu_Y^4e^{-2c_Y\sigma}+\mu_X^4+K^{T\bar{T}}D_T\hat{W}D_{\bar{T}}\hat{\bar{W}}+-3|\hat{W}|^2) \label{eq;effective},
\end{eqnarray}
where
$\hat{W}\equiv  W_{\rm{inf}}|_{X=Y=0}+ W_{\rm{min}}|_{X=Y=0}.$
The detailed discussions are given in the appendix~A. We use this effective potential in the following analyses.
\subsection{KKLT type}
We consider the model which contains the following superpotential terms $W_{\rm{min}}$ governing the whole dynamics around the minimum:
\begin{eqnarray*}
W_{\rm{min}} =w_0+A e ^{aT}+\mu ^2 _XX.\label{eq;Wmin1}
\end{eqnarray*}
Then we choose the following set of parameters:\footnote{The facters $(0.9)^n$ in Eq.~(\ref{eq;parameter}) represent a rescaling to fit the WMAP normalization at the e-foldings $N\sim50$ before the end of inflation.} 
\begin{eqnarray}
&&C=(0.9)^2\times 3\times 10^{-5}, \quad D=(0.9)^2\times 1\times 10^{-5}, \quad c=\frac{\pi}{15}, \quad d=\frac{\pi}{25}, \quad c_Y=\frac{\pi}{70},\nonumber\\
&&w_0=(0.9)^2\times 2\times 10 ^{-11}, \quad A=(0.9)^2\times 2\times 10^{-19}, \quad a=\frac{\pi}{60},\nonumber\\
&&\mu_Y =(0.9)\times 1.562633\times 10^{-3} \quad \mu _X=(0.9) \times 6.18\cdots \times 10^{-6}\label{eq;parameter},
\end{eqnarray} 
and we choose the initial condition:
\begin{eqnarray*}
\sigma(0)=19.2, \quad \sigma'(0)=0.
\end{eqnarray*}
The smallness of the parameters $C$ and $D$ can be naturally realized if moduli mixings occurs as $C=C' e^{-c_SS},D=D'e^{-d_SS}$
and $S$ is stabilized at a scale higher than the inflation scale. Therefore we don't consider the ``smallness" of parameters as a consequence of fine-tunings.
We set the parameter $\mu_X$ in such a way that the AdS minimum is uplifted to the Minkowski minimum. We just admit a fine-tuning of the parameters $\mu_X$ and $\mu_Y$ which originates from the cosmological constant problem and the generation of the inflationary inflection point\footnote{We can find inflection point inflation in string theory e.g.~\cite{Linde:2007jn}, \cite{Badziak:2008gv} and \cite{Baumann:2007np}, and in MSSM inflation models~\cite{Allahverdi:2006iq}.}, respectively. Solutions of these fine-tuning problems are beyond the scope of this paper. Aside from these deep problems, we don't need fine-tuned parameters to separate inflation scale and SUSY breaking scale as we have shown in the previous section. 
\begin{figure}[htbp]
\begin{minipage}{\textwidth}
\begin{minipage}{0.5\textwidth}
(a)\\
\centering \includegraphics[width=\textwidth]{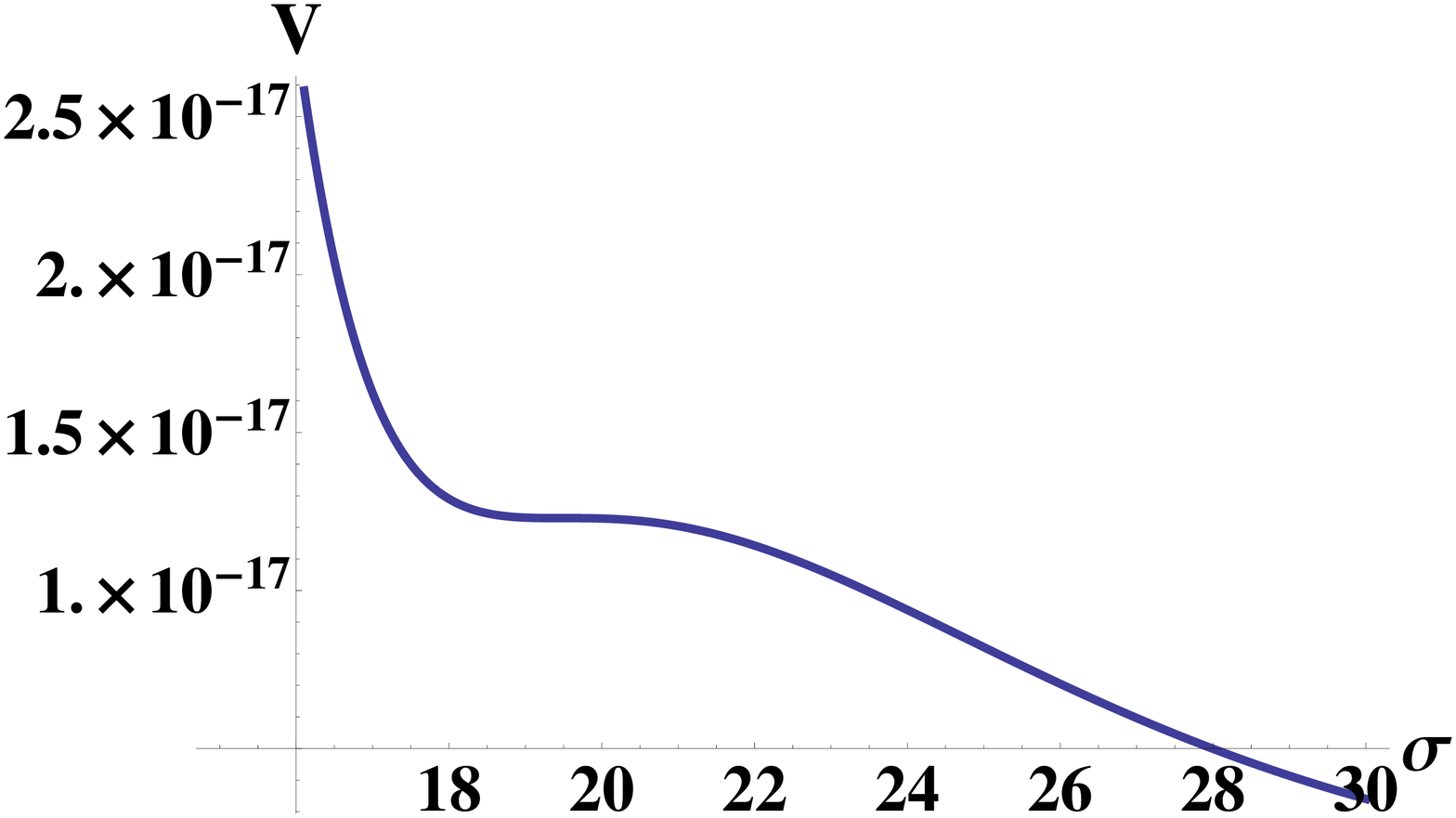}
\end{minipage}
\begin{minipage}{0.5\textwidth}
(b)\\
\centering \includegraphics[width=\textwidth]{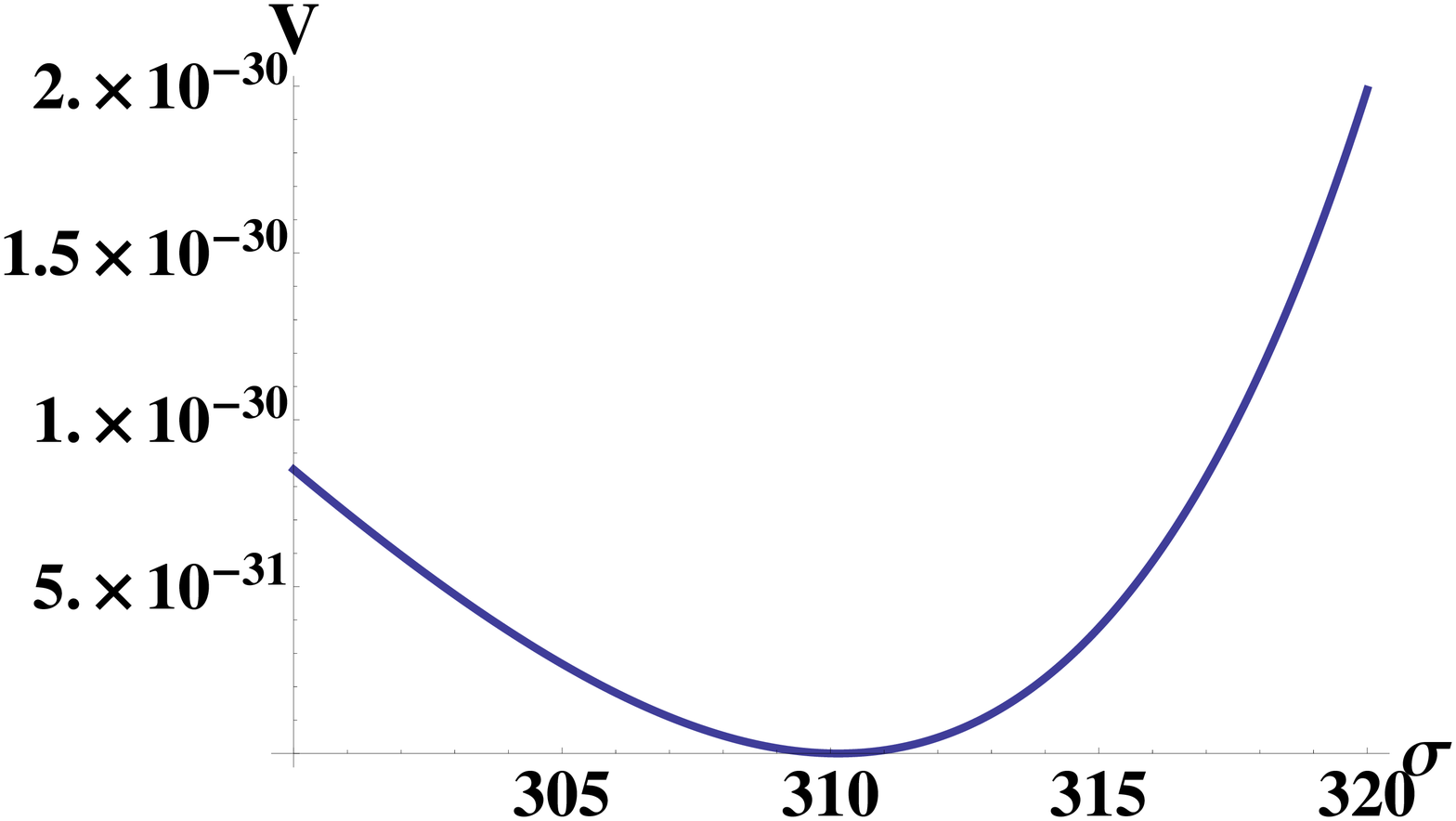}
\end{minipage}
\caption{The scalar potential for the KKLT-type model (a) in the vicinity of the inflationary inflection point and (b) in the vicinity of the minimum.}
\label{fig;potential}
\end{minipage}
\end{figure}

\begin{figure}[htbp]
\begin{center}
\includegraphics[width=0.8\textwidth]{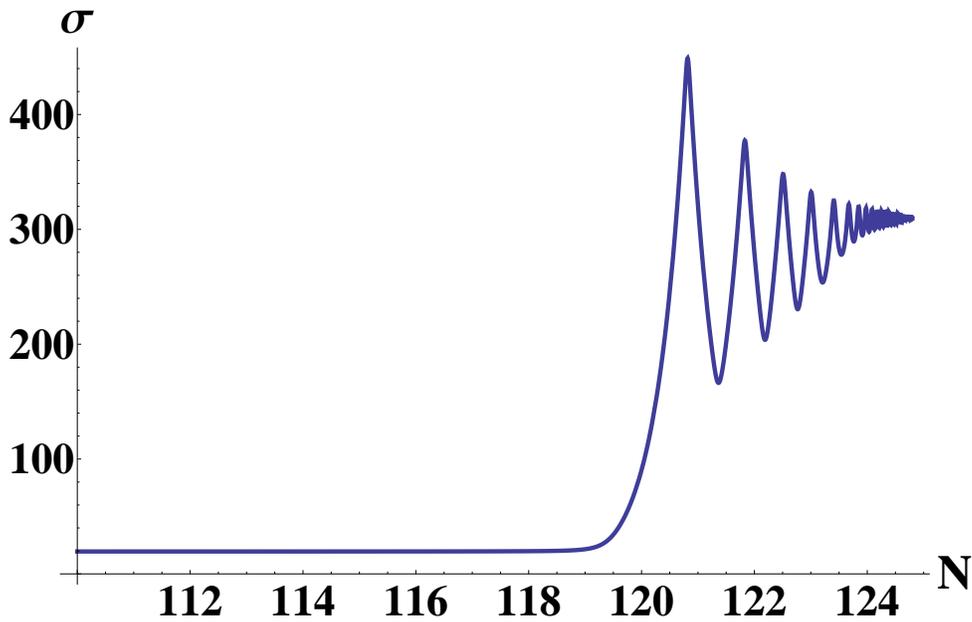}
\caption{The evolution of the inflaton $\sigma=$Re$T$ for the KKLT-type model as functions of the e-folding number $N$ in the last stage of inflation.}
\label{fig;traj}
\end{center}
\end{figure}

\begin{figure}[htbp]
\begin{minipage}{\textwidth}
\begin{minipage}{0.5\textwidth}
(a)\\
\centering\includegraphics[width=0.9\textwidth]{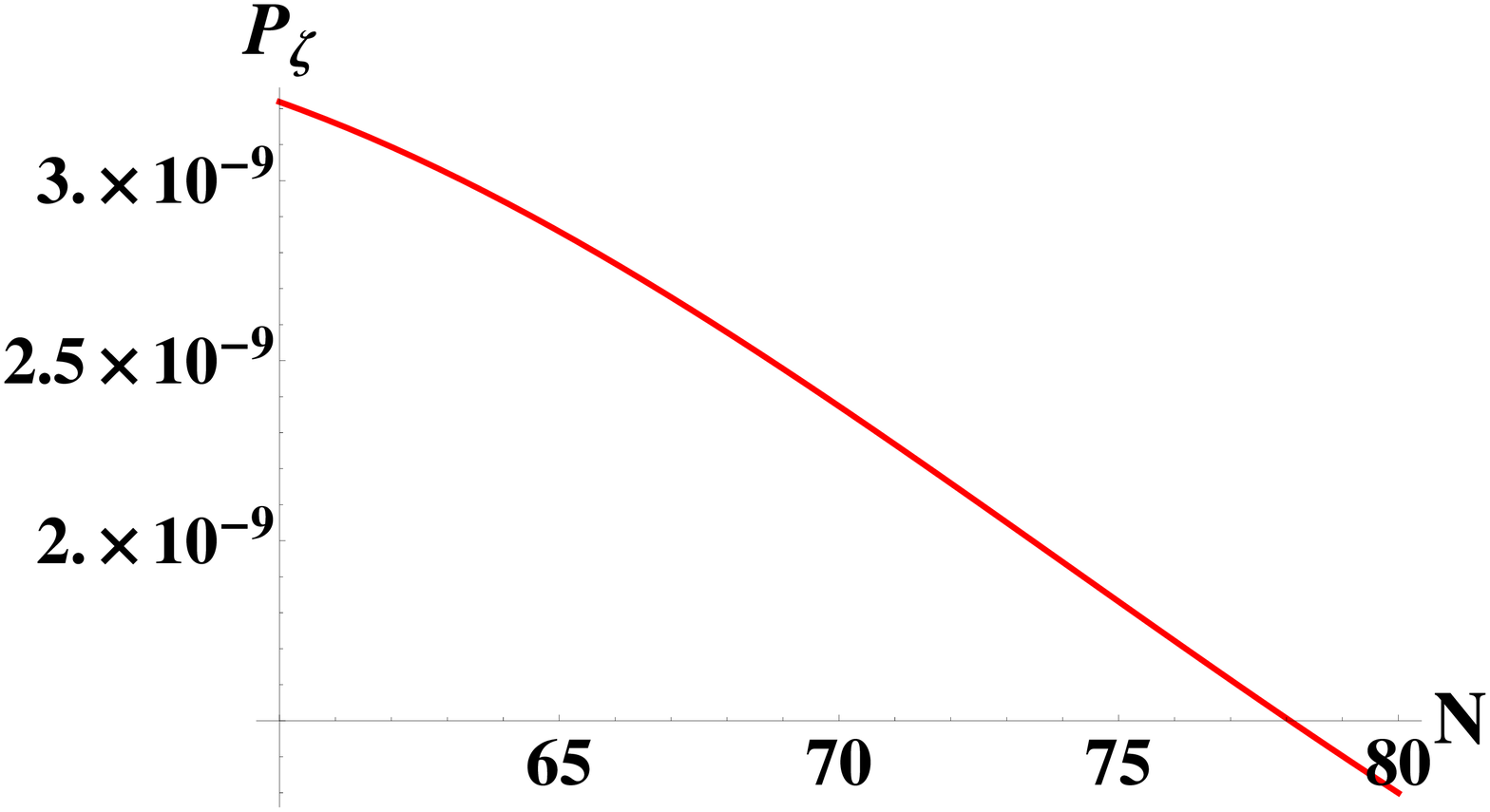}
\end{minipage}
\begin{minipage}{0.5\textwidth}
(b)\\
\centering\includegraphics[width=0.9\textwidth]{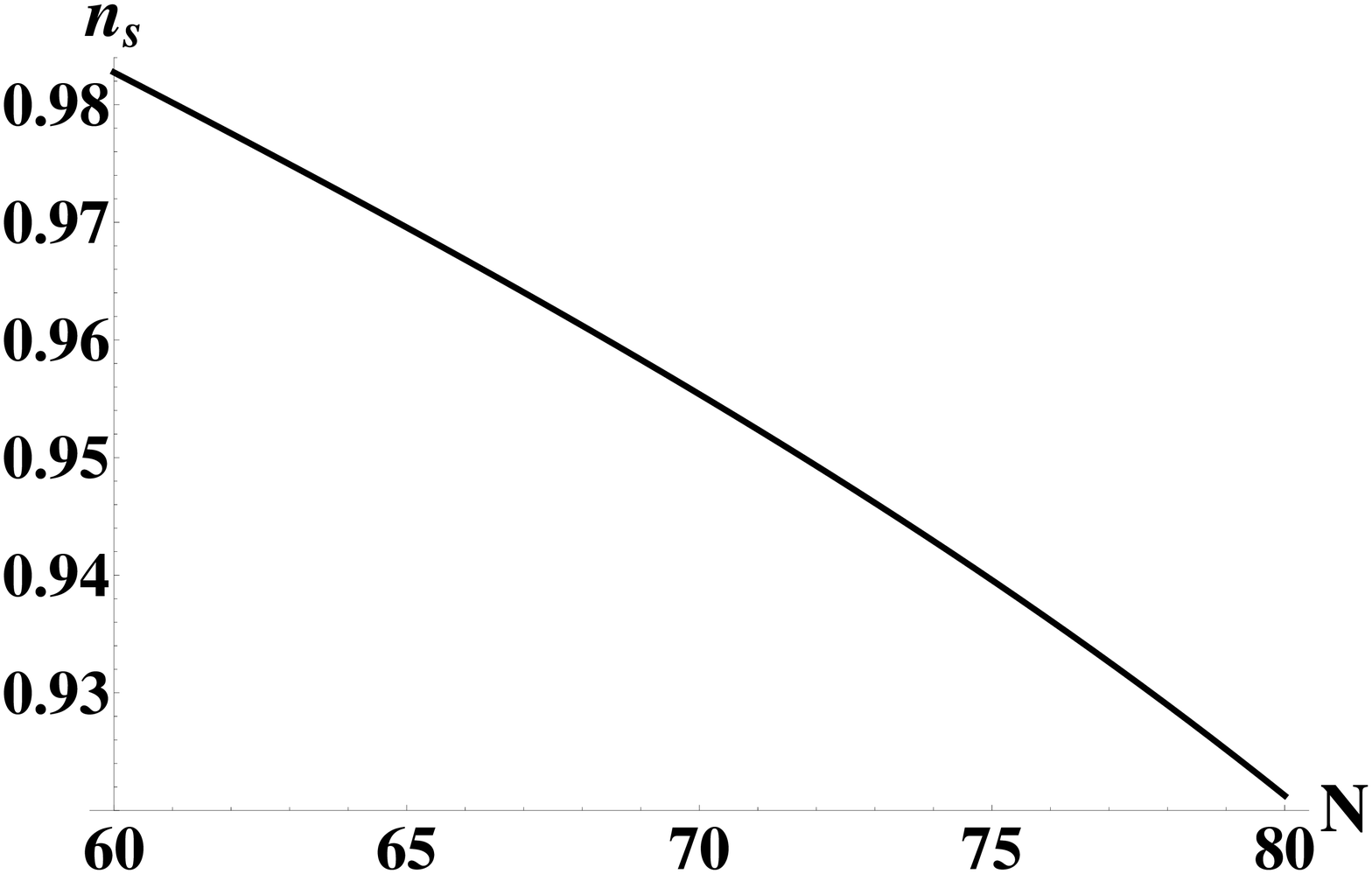}
\end{minipage}
\caption{(a) The scalar power spectrum and (b) the spectral index of the scalar power spectrum for the KKLT type model as a function of the e-folding number. The region from 40 to 60 e-foldings before the end of the inflation is focused here. }
\label{fig;obs}
\end{minipage}
\end{figure}

We show the shape of the potential around $T_{\rm{inf}}$ and the one around $T_{\rm{min}}$ in Fig.~\ref{fig;potential}. We can find that the hight of the potential is extremely suppressed  around the minimum compared with the one around the instantly uplifted de Sitter region $T\sim T_{\rm{inf}}$. The evolution of $\sigma$ is shown in Fig.~\ref{fig;traj}, and we find that the overshooting problem does not occur due to a positive exponent term. The power spectrum of the scalar curvature perturbation $\displaystyle \mathcal{P}_{\zeta}=\frac{V}{24\pi^2 \epsilon}$ and its spectral index $n_s=1+2\eta -6\epsilon$ in this model~\footnote{Because this model is treated as single field inflation with a non-canonical kinetic term generated by the K\"ahler potential (\ref{eq;kahler}), the slow roll parameters $\epsilon$ and $\eta$ are given by $\epsilon=\frac{\sigma^2(\partial_{\sigma}V)^2}{3V^2}$, $\eta=\frac{2\sigma^2\partial_{\sigma}^2V}{3V}$.}  can be found in Fig.~\ref{fig;obs}. The tensor to scalar ratio $r=16\epsilon$ in this model (and in the following two models discussed in Sec.4.2 and Sec.4.3) is $\mathcal{O}(10^{-10})$, therefore these observables $\mathcal{P}_{\zeta},n_s$ and $r$ are consistent with the CMB observation~\cite{Komatsu:2010fb}.

We derive the SUSY breaking parameters around the minimum as follows:
\begin{eqnarray*}
m_{3/2}=2.8 \ [{\rm{TeV}}], \quad F^X=4.8 [{\rm{TeV}}], \quad \sqrt{K_{T\bar{T}}}F^T=475 [{\rm{GeV}}], \quad F^Y=277 [{\rm{GeV}}].
\end{eqnarray*}
The gravitino mass is $\mathcal{O}(10^3$GeV) at the minimum, even though the Hubble parameter is $\mathcal{O}(10^9$GeV) during inflation. The hierarchy of  $\mathcal{O}(10^{6})$ is realized between the SUSY breaking scale and the inflationary one which confirms the success of the separation mechanism proposed in this paper.
\subsection{Racetrack type}
Next we adopt the racetrack-type superpotential terms
\begin{eqnarray}
W_{\rm{min}}=A e^{aT}+Be^{-bT}+\mu ^2_XX.\label{eq;Wmin2}
\end{eqnarray}
We choose the set of parameters,
\begin{eqnarray*}
&&C=(0.9)^2\times 3\times 10^{-5}, \quad D=(0.9)^2\times 1\times 10^{-5}, \quad c=\frac{\pi}{15}, \quad d=\frac{\pi}{25}, \quad c_Y=\frac{\pi}{70},\\
&&A=(0.9)^2\times 5\times10^{-14}, \quad B=(0.9)^2\times 6\times 10^{-9}, \quad a=\frac{\pi}{175},\quad b=\frac{\pi}{140},\\
&&\mu_Y =(0.9)\times 1.562633\times 10^{-3}, \quad \mu_X=(0.9)\times5.66\cdots \times10^{-6},
\end{eqnarray*}
and the following initial conditions,
\begin{eqnarray*}
\sigma(0)=19.4, \quad \sigma'(0)=0.
\end{eqnarray*}
These parameters in $W_{\rm{inf}}$ are almost same as those in the subsection 4.1. So, the evolution of the modulus, the spectral index, the power spectrum, and the tensor to scalar ratio in this model are similar to those in the previous subsection.  SUSY breaking parameters in this model is found as
\begin{eqnarray*}
&&m_{3/2}=2.4 \ [{\rm{TeV}}], \quad F^X=4.1 [{\rm{TeV}}],\\
&&\sqrt{K_{T\bar{T}}}F^T=291 [{\rm{GeV}}], \quad F^Y=295 [{\rm{GeV}}].
\end{eqnarray*}
Again low scale SUSY breaking is realized and the separation is successful.
\subsection{R-symmetric type}
Finally, we consider the following superpotential terms:
\begin{eqnarray*}
W_{\rm{min}}=Ae^{aT}+\mu^2_XX.\label{eq;Wmin3}
\end{eqnarray*}
Then, we choose the following set of parameters,
\begin{eqnarray*}
&&C=3\times 10^{-5}, \quad D=1\times 10^{-5}, \quad c=\frac{\pi}{15}, \quad d=\frac{\pi}{25}, \quad c_Y=\frac{\pi}{70},\\
&&A=2\times10^{-12}, \quad a=\frac{\pi}{370}, \quad \mu_Y=1.562651 \times 10^{-6}, \quad \mu_X=5.63\cdots \times 10^{-6},
\end{eqnarray*}
and the initial conditions,
\begin{eqnarray*}
\sigma(0)=19.45, \quad \sigma'(0)=0.
\end{eqnarray*}
The observables during inflation are similar to the previous models, however, SUSY breaking parameters differ substantially from the other ones. We show the SUSY breaking parameters in this model:
\begin{eqnarray*}
&&m_{3/2}=4.3 \ [{\rm{TeV}}], \quad F^X=5.0 [{\rm{TeV}}],\\
&&\sqrt{K_{T\bar{T}}}F^T=5.4 [{\rm{TeV}}], \quad F^Y=430 [{\rm{GeV}}].
\end{eqnarray*}
We find that the F-term of the modulus is comparable with that of the SUSY breaking sector $X$ and the gravitino mass. The difference from the previous models is caused by the existence of  R-symmetry. In this model, the superpotential $W_{\rm{min}}$ has an R-symmetry, and the effective potential around the minimum has an approximate R-symmetry. In Ref.~\cite{Abe:2007ax}, it is pointed out that there is an R-symmetric SUSY breaking minimum for the modulus whose imaginary part is shifted under the R-symmetry transformation\footnote{We would like to thank Hiroyuki Abe for noticing this point.}. Therefore, the modulus in this model also plays a SUSY breaking field, and the F-term of the modulus becomes relatively large. Such SUSY breaking parameters produce a different pattern of superpaticle spectrum. That is relevant to particle phenomenology.
\subsection{Comments on the difference between the instant uplifted inflation and the KL model}
We give some comments on the difference between the instant uplifted inflation and the KL model. The KL model is attractive from the viewpoint of its simplicity, and the phenomenological consequences of the KL stabilization is studied recently. As shown in the Ref. \cite{Linde:2011ja}, the volume modulus gets a heavy SUSY mass, and then the F-term of the modulus stabilized by the KL type superpotential\ref{KL} is much smaller than that of the original KKLT volume modulus, then the contribution from such a F-term is much smaller than the anomaly mediation. Therefore the gaugino masses are dominated by the anomaly mediation, and the gravitino mass $m_{3/2}$ have to be $\mathcal{O}(100)$TeV to get the sufficient gluino mass. On the other hand, MSSM sfermions get masses of order $m_{3/2}$ because they are not sequestered to avoid the tachyonic mass problem. 

In our models, the F-term of the volume modulus is relatively large, and such F-terms can play an important role from the phenomenological viewpoint. For example, in the original KKLT type models (shown in Sec.~4.1 or 4.2), the anomaly mediation and the modulus mediation can be comparable, and then we can realize the mirage mediation \cite{Choi:2005uz} which is an attractive scenario because it solves the little hierarchy problem elegantly \cite{Choi:2005hd}. In our model, the F-term of the volume modulus $T$ is determined by the $W_{\rm{min}}$ and then we can obtain various patterns of phenomenological consequences. This feature is an important difference between the KL model and our model.
\section{Conclusion}
We proposed a new class of a mechanism to separate the scale of inflation with KKLT type moduli stabilization from SUSY breaking scale.  The two ingredients are required to achieve the separation. One is the existence of ``uplifton" which has a following form of F-term $F^Y\sim\mu_Y^2e^{-c_YT}$, then an inflationary de Sitter point can be realized. Because the F-term of uplifton decreases exponentially as $T$ increases,  we could make the minimum where the scale of the scalar potential is extremely smaller than the one during inflation. The other ingredient is a positive exponent term in superpotential like $\tilde{C}e^{cT}$, which prevent the overshooting after inflation. As we have shown, we don't need fine-tuned parameters to separate the two scales. Due to the separation mechanism, we could adopt some different patterns of stabilization potential after inflation. Therefore, different phenomenological models can be realized. For example, KKLT type stabilization in our model may realize the mirage mediation \cite{Choi:2005uz}, \cite{Choi:2005hd} which solves the SUSY little hierarchy problem.

 In this paper, we focused on the separation between inflation and SUSY breaking scale.
In order to construct realistic models, we have to combine these models with the successful Big-Bang nucleosynthesis. In addition, low scale SUSY breaking models predict SUSY particles with TeV scale masses, and they may be discovered at the LHC in the near future. So, it is also important to analyze predictions of such SUSY models. We will investigate concrete phenomenological models combined with the instant uplifted inflation proposed in this paper as a future work~\cite{prog}. 

\section*{Acknowledgments}
The author especially would like to thank Hiroyuki Abe for the early collaboration, useful discussions, and reading the manuscript carefully, and Tetsutaro Higaki for many useful comments and discussions. He is also grateful to Hajime Otsuka, Keigo Sumita, Yoshiyuki Tatsuta for discussions and comments.
\appendix
\renewcommand{\theequation}{\Alph{section}.\arabic{equation}}
\setcounter{equation}{0}
\section{Derivation of the effective potential}
The K$\ddot{\rm{a}}$hler potential and superpotential in our model are given as follows:
\begin{eqnarray}
K&=&-3\log (T+\bar{T})+|X|^2-\frac{|X|^4}{\Lambda ^2}+|Y|^2-\frac{|Y|^4}{\Lambda'^2},\\
W&=&\mu_X^2X+\mu_Y^2Ye^{-c_YT}+\hat{W},\\
\hat{W}&=&W_{\rm{inf}}|_{X=Y=0}+W_{\rm{min}}|_{X=Y=0},
\end{eqnarray}
where $\Lambda,\Lambda'\ll 1$, $W_{\rm{inf}}$ is given by Eq.~(\ref{eq;Winf}) and $W_{\rm{min}}$ takes some patterns given by Eq.~(\ref{eq;Wmin1}), Eq.~(\ref{eq;Wmin2}), and Eq.~(\ref{eq;Wmin3}).

The F-term scalar potential is generically given by
\begin{eqnarray}
V=e^K(D_IWK^{I\bar{J}}D_{\bar{J}}\bar{W}-3|W|^2).
\end{eqnarray}
Then, we expand the potential in powers of $X$ and $Y$, up to their quadratic terms. The terms of the 0th, 1st, and 2nd order of $X$ and $Y$ are represented respectively as $V^{(0)},V^{(1)},V^{(2)}$:
\begin{eqnarray}
V^{(0)}&=&\frac{1}{8\sigma^3}(\mu_X^4+\mu_Y^4e^{-2c_Y\sigma}+D_T\hat{W}K^{T\bar{T}}D_{\bar{T}}\hat{\bar{W}}-3|\hat{W}|^2)\nonumber \\
&\equiv &\frac{1}{8\sigma^3}V_0,\\
\nonumber\\
V^{(1)}&=&\frac{1}{8\sigma^3}(K_TK^{T\bar{T}}D_{\bar{T}}\hat{\bar{W}}-2\hat{\bar{W}})\mu_X^2 X \nonumber\\
&&+\frac{1}{8\sigma^3}(K^{T\bar{T}}D_{\bar{T}}\hat{W}(K_T-c_Y)-2\hat{\bar{W}})\mu_Y^2e^{-c_Y T}Y+\rm{h.c.},\\
\nonumber\\
V^{(2)}&=&\frac{1}{8\sigma^3}\left[ V_0+\frac{4\mu_X^4}{\Lambda^2}+2\mu_X^2+|\hat{W}|^2\right] |X|^2\nonumber\\
&&+\frac{1}{8\sigma ^3}\left[ V_0+\frac{4\mu_Y^4e^{-2c_Y\sigma}}{\Lambda'^2}+(K^{T\bar{T}}(K_T-c_Y)(K_{\bar{T}}-c_Y)-1)\mu_Y^4 e^{-2c_Y\sigma} +|\hat{W}|^2\right] |Y|^2\nonumber\\
&&+\frac{1}{4\sigma^3}(1+c_Y\sigma)\mu_X^2 \mu_Y^2 e^{-c_Y\bar{T}}X\bar{Y}+\frac{1}{4\sigma^3}(1+c_Y\sigma)\mu_X^2 \mu_Y^2 e^{-c_YT}Y\bar{X}.\nonumber\\
\end{eqnarray}
The extremum conditions in terms of $\bar{X}$ and $\bar{Y}$ are found respectively as
\begin{eqnarray}
&&\left[V_0+\frac{4\mu_X^4}{\Lambda^2}+2\mu_X^2+|\hat{W}|^2\right] X+2(1+c_Y\sigma)\mu_X^2\mu_Y^2e^{-c_YT}Y=2(\sigma D_T\hat{W}+\hat{W})\mu_X^2,\nonumber\\
 \label{eq;X}\\[15pt]
&&\left[V_0+\frac{4\mu_Y^4e^{-2c_Y\sigma}}{\Lambda'^2}+(K^{T\bar{T}}(K_T-c_Y)(K_{\bar{T}}-c_Y)-1)\mu_Y^4 e^{-2c_Y\sigma} +|\hat{W}|^2\right] \nonumber\\
&&+2(1+c_Y\sigma)\mu_X^2\mu_Y^2e^{-c_Y\bar{T}}X=(2\hat{\bar{W}}-K^{T\bar{T}}D_{\bar{T}}\hat{W}(K_{\bar{T}}-c_Y))\mu_Y^2e^{-c_Y \bar{T}}. \label{eq;Y}
\end{eqnarray}
For a notational convenience, we define the following quantities:
\begin{eqnarray}
V_{Y\bar{Y}}&\equiv& \left[V_0+\frac{4\mu_Y^4e^{-2c_Y\sigma}}{\Lambda'^2}+(K^{T\bar{T}}(K_T-c_Y)(K_{\bar{T}}-c_Y)-1)\mu_Y^4 e^{-2c_Y\sigma} +|\hat{W}|^2\right],\nonumber\\
\\
V_{X\bar{X}}&\equiv& \left[V_0+\frac{4\mu_X^4}{\Lambda^2}+2\mu_X^2+|\hat{W}|^2\right], \\
V_{X\bar{Y}}&\equiv& 2(1+c_Y\sigma)\mu_X^2\mu_Y^2e^{-c_Y\bar{T}},\\
V_{Y\bar{X}}&\equiv& 2(1+c_Y\sigma)\mu_X^2\mu_Y^2e^{-c_YT},\\
V_{\bar{X}}|_0&\equiv& 2(\sigma D_T\hat{W}+\hat{W})\mu_X^2,\\
V_{\bar{Y}}|_0&\equiv& (2\hat{\bar{W}}-K^{T\bar{T}}D_{\bar{T}}\hat{W}(K_{\bar{T}}-c_Y))\mu_Y^2e^{-c_Y \bar{T}}.
\end{eqnarray}
Using these notations, Eqs.~(\ref{eq;X}), (\ref{eq;Y}) are represented by
\begin{eqnarray}
\left( \begin{array}{cc} V_{X\bar{X}} &V_{Y\bar{X}} \\ V_{X\bar{Y}} & V_{Y\bar{Y}}  \end{array} \right) \left( \begin{array}{cc} X\\ Y \end{array} \right) &= &\left( \begin{array}{cc} V_{\bar{X}}|_0\\ V_{\bar{Y}}|_0 \end{array} \right),
\end{eqnarray}
that can be rewritten as 
\begin{eqnarray}
 \left( \begin{array}{cc} X\\ Y \end{array} \right) &\sim & \frac{1}{V_{X\bar{X}}V_{Y\bar{Y}}}\left( \begin{array}{cc} V_{Y\bar{Y}}&-V_{Y\bar{X}}\\ -V_{X\bar{Y}}&V_{X\bar{X}} \end{array} \right)\left( \begin{array}{cc} V_{\bar{X}}|_0\\ V_{\bar{Y}}|_0\end{array}\right)\\[10pt]
& =&\left( \begin{array}{cc} \displaystyle \frac{V_{\bar{X}}|_0}{V_{X\bar{X}}}-\frac{V_{\bar{Y}}|_0 V_{Y\bar{X}}}{V_{X\bar{X}}V_{Y\bar{Y}}}\\[15pt] \displaystyle \frac{V_{\bar{Y}}|_0}{V_{Y\bar{Y}}}-\frac{V_{\bar{X}}|_0 V_{X\bar{Y}}}{V_{X\bar{X}}V_{Y\bar{Y}}} \end{array}\right).
\end{eqnarray}
Then, we can evaluate the VEV of $X$ as follows:
\begin{eqnarray}
X\sim &&\frac{2\mu_X^2( \sigma D_T\hat{W}+\hat{W})}{\left[V_0+\frac{4\mu_X^4}{\Lambda^2}+2\mu_X^2+|\hat{W}|^2\right]}
-\frac{2\mu_X^2\mu_Y^4e^{-2c_Y\sigma}(1+c_Y\sigma) \{ 2\hat{W}-K^{T\bar{T}}D_T\hat{W}(K_{\bar{T}}-c_Y)\}}{\left[V_0+\frac{4\mu_X^4}{\Lambda^2}+|\hat{W}|^2\right]\left[V_0+\frac{4\mu_Y^4}{\Lambda'^2}e^{-2c_Y\sigma}+|\hat{W}|^2\right]}\notag\\
\\[10pt]
&&\leq \frac{2\mu_X^2( \sigma D_T\hat{W}+\hat{W})}{\left[V_0+\frac{4\mu_X^4}{\Lambda^2}+2\mu_X^2+|\hat{W}|^2\right]}
-\Lambda '^2\frac{2\mu_X^2 (1+c_Y\sigma) \{ 2\hat{W}-K^{T\bar{T}}D_T\hat{W}(K_{\bar{T}}-c_Y)\}}{\left[V_0+\dfrac{4\mu_X^4}{\Lambda^2}+|\hat{W}|^2\right]}.\nonumber\\
\end{eqnarray}
We take account the relation $\mathcal{O}(\sigma D_T\hat{W})\sim \mathcal{O}(\hat{W})\sim\mathcal{O}(\sqrt{V_0})$. In the case $\mathcal{O}(\sqrt{V_0})\geq \mathcal{O}\left(\frac{\mu_X^2}{\Lambda}\right)$, we find 
\begin{eqnarray}
X\leq \mathcal{O}\left(\frac{\mu_X^2}{\sqrt{V_0}}\right)\leq \mathcal{O}(\Lambda)\ll 1.
\end{eqnarray}
On the other hand, in the case $\mathcal{O}(\sqrt{V_0})\leq \mathcal{O}\left(\frac{\mu_X^2}{\Lambda}\right)$, we find 
\begin{eqnarray}
X\leq \mathcal{O}\left(\frac{\mu_X^2 \sqrt{V_0}}{(\frac{\mu_X^2}{\Lambda})^2}\right)\leq \mathcal{O}(\Lambda)\ll 1.
\end{eqnarray}
As a result, we can always derive the relation $X\leq\Lambda$. From similar discussions, we can find the relation $Y\leq\Lambda'$ . These relations show that we can neglect the VEVs $ X$ and $Y$, and fluctuations of these fields around the VEVs have a large mass during inflation. Therefore we neglect the small VEVs and the fluctuations of $X$ and $Y$, and find the effective potential $V_{\rm{eff}}$ for $T$ shown in Eq.~(\ref{eq;effective}).


\begin{thebibliography}{99}
\bibitem{Guth:1980zm} 
  A.~H.~Guth,
  Phys.\ Rev.\ D {\bf 23}, 347 (1981);
  
  A.~A.~Starobinsky,
  Phys.\ Lett.\ B {\bf 91}, 99 (1980);
  
  K.~Sato,
  Mon.\ Not.\ Roy.\ Astron.\ Soc.\  {\bf 195}, 467 (1981).
  
\bibitem{Mazumdar:2010sa} 
  A.~Mazumdar and J.~Rocher,
  Phys.\ Rept.\  {\bf 497}, 85 (2011)
  [arXiv:1001.0993 [hep-ph]].
  
\bibitem{Kachru:2003aw} 
  S.~Kachru, R.~Kallosh, A.~D.~Linde and S.~P.~Trivedi,
  Phys.\ Rev.\ D {\bf 68}, 046005 (2003)
  [hep-th/0301240].
  
\bibitem{Kallosh:2004yh} 
  R.~Kallosh and A.~D.~Linde,
  JHEP {\bf 0412}, 004 (2004)
  [hep-th/0411011].
  
\bibitem{BlancoPillado:2004ns} 
  J.~J.~Blanco-Pillado, C.~P.~Burgess, J.~M.~Cline, C.~Escoda, M.~Gomez-Reino, R.~Kallosh, A.~D.~Linde and F.~Quevedo,
  JHEP {\bf 0411}, 063 (2004)
  [hep-th/0406230].
  
\bibitem{Linde:2007jn} 
  A.~D.~Linde and A.~Westphal,
  JCAP {\bf 0803}, 005 (2008)
  [arXiv:0712.1610 [hep-th]].
  
\bibitem{Conlon:2005jm} 
  J.~P.~Conlon and F.~Quevedo,
  JHEP {\bf 0601}, 146 (2006)
  [hep-th/0509012].
  
\bibitem{Badziak:2008yg} 
  M.~Badziak and M.~Olechowski,
  JCAP {\bf 0807}, 021 (2008)
  [arXiv:0802.1014 [hep-th]].
  
\bibitem{Covi:2008cn} 
  L.~Covi, M.~Gomez-Reino, C.~Gross, J.~Louis, G.~A.~Palma and C.~A.~Scrucca,
  JHEP {\bf 0808}, 055 (2008)
  [arXiv:0805.3290 [hep-th]].
  
\bibitem{Badziak:2008gv} 
  M.~Badziak and M.~Olechowski,
  JCAP {\bf 0902}, 010 (2009)
  [arXiv:0810.4251 [hep-th]].
  
\bibitem{Abe:2005rx} 
  H.~Abe, T.~Higaki and T.~Kobayashi,
  Phys.\ Rev.\ D {\bf 73}, 046005 (2006)
  [hep-th/0511160];
  
  H.~Abe, T.~Higaki and T.~Kobayashi,
  Nucl.\ Phys.\ B {\bf 742}, 187 (2006)
  [hep-th/0512232];
 
  H.~Abe, T.~Higaki, T.~Kobayashi and O.~Seto,
  Phys.\ Rev.\ D {\bf 78}, 025007 (2008)
  [arXiv:0804.3229 [hep-th]].
  
\bibitem{Abe:2006xp} 
 E.~Dudas, C.~Papineau and S.~Pokorski,
  JHEP {\bf 0702}, 028 (2007)
  [hep-th/0610297];
  
  H.~Abe, T.~Higaki, T.~Kobayashi and Y.~Omura,
  Phys.\ Rev.\ D {\bf 75}, 025019 (2007)
  [hep-th/0611024];
  
  H.~Abe, T.~Higaki and T.~Kobayashi,
  Phys.\ Rev.\ D {\bf 76}, 105003 (2007)
  [arXiv:0707.2671 [hep-th]].
  
\bibitem{Kobayashi:2010rx} 
  T.~Kobayashi and M.~Sakai,
  JHEP {\bf 1104}, 121 (2011)
  [arXiv:1012.2187 [hep-th]].

\bibitem{He:2010uk} 
  T.~He, S.~Kachru and A.~Westphal,
  JHEP {\bf 1006}, 065 (2010)
  [arXiv:1003.4265 [hep-th]];

  S.~Antusch, K.~Dutta and S.~Halter,
  JHEP {\bf 1203}, 105 (2012)
  [arXiv:1112.4488 [hep-th]].
  
\bibitem{Kallosh:2011qk} 
  R.~Kallosh, A.~Linde, K.~A.~Olive and T.~Rube,
  Phys.\ Rev.\ D {\bf 84}, 083519 (2011)
  [arXiv:1106.6025 [hep-th]].
  
\bibitem{Linde:2011ja} 
  A.~Linde, Y.~Mambrini and K.~A.~Olive,
  Phys.\ Rev.\ D {\bf 85}, 066005 (2012)
  [arXiv:1111.1465 [hep-th]].

  E.~Dudas, A.~Linde, Y.~Mambrini, A.~Mustafayev and K.~A.~Olive,
  Eur.\ Phys.\ J.\ C {\bf 73}, 2268 (2013)
  [arXiv:1209.0499 [hep-ph]].
\bibitem{Abe:2006xi} 
  H.~Abe, T.~Higaki and T.~Kobayashi,
  Phys.\ Rev.\ D {\bf 74}, 045012 (2006)
  [hep-th/0606095].
 
\bibitem{Badziak:2009eh} 
  M.~Badziak and M.~Olechowski,
  JCAP {\bf 1002}, 026 (2010)
  [arXiv:0911.1213 [hep-th]].
 
\bibitem{Conlon:2008cj} 
  J.~P.~Conlon, R.~Kallosh, A.~D.~Linde and F.~Quevedo,
  JCAP {\bf 0809}, 011 (2008)
  [arXiv:0806.0809 [hep-th]].

\bibitem{Florea:2006si} 
  B.~Florea, S.~Kachru, J.~McGreevy and N.~Saulina,
  JHEP {\bf 0705}, 024 (2007)
  [hep-th/0610003];
  
 R.~Blumenhagen, M.~Cvetic and T.~Weigand,
  Nucl.\ Phys.\ B {\bf 771}, 113 (2007)
  [hep-th/0609191];

  N.~Akerblom, R.~Blumenhagen, D.~Lust and M.~Schmidt-Sommerfeld,
  JHEP {\bf 0708}, 044 (2007)
  [arXiv:0705.2366 [hep-th]].

\bibitem{Cvetic:2008mh} 
  M.~Cvetic and T.~Weigand,
  arXiv:0807.3953 [hep-th];
  
J.~J.~Heckman, J.~Marsano, N.~Saulina, S.~Schafer-Nameki and C.~Vafa,
  arXiv:0808.1286 [hep-th];
  
 E.~Dudas, Y.~Mambrini, S.~Pokorski, A.~Romagnoni and M.~Trapletti,
  JHEP {\bf 0903}, 011 (2009)
  [arXiv:0809.5064 [hep-th]];
  
  P.~G.~Camara, C.~Condeescu, E.~Dudas and M.~Lennek,
  JHEP {\bf 1006}, 062 (2010)
  [arXiv:1003.5805 [hep-th]].
  
\bibitem{Abe:2006eg} 
  H.~Abe and Y.~Sakamura,
  Phys.\ Rev.\ D {\bf 75}, 025018 (2007)
  [hep-th/0610234].
 
\bibitem{Lukas:1997fg} 
  A.~Lukas, B.~A.~Ovrut and D.~Waldram,
  Nucl.\ Phys.\ B {\bf 532}, 43 (1998)
  [hep-th/9710208];.
  
  A.~Lukas, B.~A.~Ovrut and D.~Waldram,
  Phys.\ Rev.\ D {\bf 57}, 7529 (1998)
  [hep-th/9711197];
  
  E.~I.~Buchbinder and B.~A.~Ovrut,
  Phys.\ Rev.\ D {\bf 69}, 086010 (2004)
  [hep-th/0310112].

\bibitem{Cascales:2003zp} 
  J.~F.~G.~Cascales and A.~M.~Uranga,
  JHEP {\bf 0305}, 011 (2003)
  [hep-th/0303024];
  
 F.~Marchesano and G.~Shiu,
  JHEP {\bf 0411}, 041 (2004)
  [hep-th/0409132].
  
\bibitem{Ceresole:2000jd} 
  A.~Ceresole and G.~Dall'Agata,
  Nucl.\ Phys.\ B {\bf 585}, 143 (2000)
  [hep-th/0004111].
  
\bibitem{Baumann:2007np} 
  D.~Baumann, A.~Dymarsky, I.~R.~Klebanov, L.~McAllister and P.~J.~Steinhardt,
  Phys.\ Rev.\ Lett.\  {\bf 99}, 141601 (2007)
  [arXiv:0705.3837 [hep-th]].
  
\bibitem{Allahverdi:2006iq} 
  R.~Allahverdi, K.~Enqvist, J.~Garcia-Bellido and A.~Mazumdar,
  Phys.\ Rev.\ Lett.\  {\bf 97}, 191304 (2006)
  [hep-ph/0605035];

  R.~Allahverdi, K.~Enqvist, J.~Garcia-Bellido, A.~Jokinen and A.~Mazumdar,
  JCAP {\bf 0706}, 019 (2007)
  [hep-ph/0610134];
  
   R.~Allahverdi, B.~Dutta and A.~Mazumdar,
  Phys.\ Rev.\ D {\bf 78}, 063507 (2008)
  [arXiv:0806.4557 [hep-ph]].

\bibitem{Komatsu:2010fb} 
  E.~Komatsu {\it et al.}  [WMAP Collaboration],
  Astrophys.\ J.\ Suppl.\  {\bf 192}, 18 (2011)
  [arXiv:1001.4538 [astro-ph.CO]].
  
\bibitem{Abe:2007ax} 
  H.~Abe, T.~Kobayashi and Y.~Omura,
  JHEP {\bf 0711}, 044 (2007)
  [arXiv:0708.3148 [hep-th]].
  
\bibitem{Choi:2005uz} 
  K.~Choi, K.~S.~Jeong and K.~-i.~Okumura,
  JHEP {\bf 0509}, 039 (2005)
  [hep-ph/0504037].

 M.~Endo, M.~Yamaguchi and K.~Yoshioka,
  Phys.\ Rev.\ D {\bf 72}, 015004 (2005)
  [hep-ph/0504036].
  
\bibitem{Choi:2005hd} 
  K.~Choi, K.~S.~Jeong, T.~Kobayashi and K.~-i.~Okumura,
  Phys.\ Lett.\ B {\bf 633}, 355 (2006)
  [hep-ph/0508029].

  R.~Kitano and Y.~Nomura,
  Phys.\ Lett.\ B {\bf 631}, 58 (2005)
  [hep-ph/0509039].

 K.~Choi, K.~S.~Jeong, T.~Kobayashi and K.~-i.~Okumura,
  Phys.\ Rev.\ D {\bf 75}, 095012 (2007)
  [hep-ph/0612258].
  
\bibitem{prog}
 Y.~Yamada et al. \quad in progress  
\end{thebibliography}
\end{document}